\documentclass[reprint,amssymb, amsmath, aps, showpacs, footinbib,  prb]{revtex4-1}
\usepackage{graphicx}

\newcommand{\eff}{\text{eff}}
\newcommand{\AFM}{\text{AFM}}
\newcommand{\FM}{\text{FM}}
\newcommand{\FLL}{\text{FLL}}
\newcommand{\AMF}{\text{AMF}}

\begin{document}

\title{Unusual ferromagnetic superexchange in CdVO$_3$: The role of Cd}

\author{Alexander A. Tsirlin}
\email{altsirlin@gmail.com}

\author{Oleg Janson}

\author{Helge Rosner}
\email{Helge.Rosner@cpfs.mpg.de}

\affiliation{Max Planck Institute for Chemical Physics of Solids, N\"{o}thnitzer
Str.~40, 01187 Dresden, Germany}

\begin{abstract}
A microscopic magnetic model of the low-pressure modification of CdVO$_3$ is established, based on density functional theory (DFT) band-structure calculations, magnetization measurements, and quantum Monte-Carlo simulations. This compound is a rare example of a quasi-one-dimensional spin-$\frac12$ system showing exclusively ferromagnetic exchange. The spin lattice of CdVO$_3$ entails zigzag chains with an effective intrachain coupling $J\simeq -90$~K and interchain couplings of $J_c\simeq -18$~K and $J_a\simeq -3$~K. Quantum fluctuations are partially suppressed by the sizable interchain coupling $J_c$ that leads to an intermediate regime between one-dimensional and two-dimensional ferromagnetic systems. Apart from the peculiar spin model, CdVO$_3$ features an unusual mechanism of ferromagnetic superexchange. The couplings largely originate from Cd $5s$ states mediating hoppings between half-filled and empty $3d$ states of V$^{+4}$. 
\end{abstract}

\maketitle

\section{Introduction}
\label{sec:intro}
Ferromagnetism and antiferromagnetism, the two opponent magnetic interactions, are rarely balanced, because specific mechanisms of the magnetic exchange tend to favor one of the two options. For example, itinerant systems are prone to ferromagnetic Stoner instabilities, whereas superexchange in magnetic insulators is a source of mostly antiferromagnetic (AFM) interactions.\cite{anderson} This makes AFM ground states more common among insulating transition-metal compounds. Antiferromagnetism utterly dominates in low-dimensional magnets, where long-range and typically AFM couplings between the low-dimensional units (chains or layers) induce the overall AFM order. In particular, most of the quasi-one-dimensional (1D) spin-$\frac12$ ferromagnetic-chain systems are antiferromagnetically ordered because the interchain couplings are AFM.\cite{landee1979,*dupas1982,swank1979,willett1980,*hoogerbeets1985} Low-dimensional spin-$\frac12$ magnets with ferromagnetic (FM) ground state are still rare and restricted to systems based on organic radicals\cite{takahashi1991,*nakazawa1992,shimizu2006} or Cu$^{+2}$ compounds with non-trivial orbital ordering.\cite{yamada1972,*ito1976,khomskii1973,dejongh1976,*willett1988}

\begin{figure}
\includegraphics{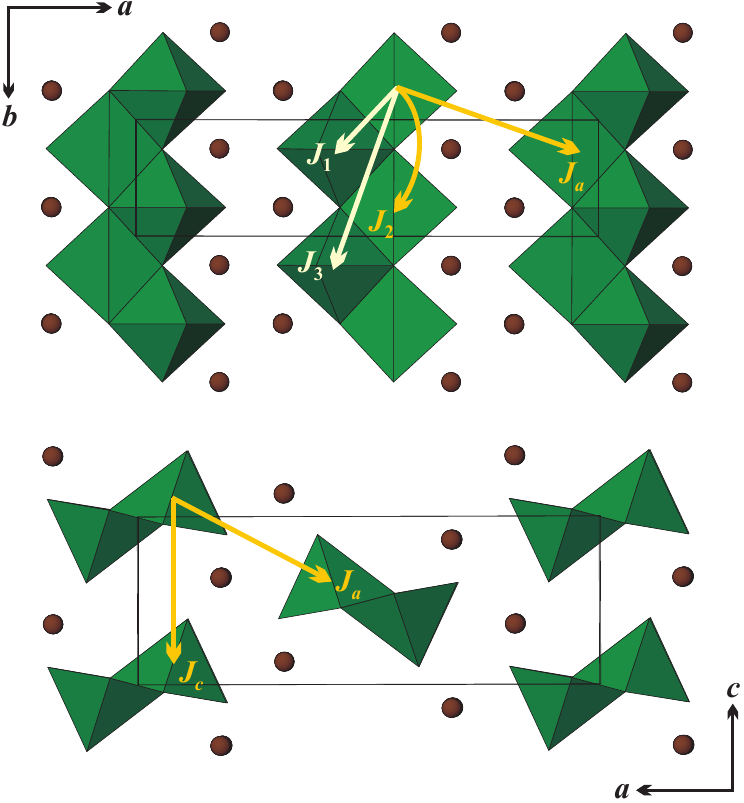}
\caption{\label{fig:structure}
(Color online) Crystal structure of CdVO$_3$ featuring zigzag chains of VO$_5$ pyramids. The chains are separated by Cd atoms (spheres). 
}
\end{figure}

The aforementioned trend is violated by one peculiar compound, the low-pressure modification%
\footnote{Throughout the paper, we refer to the low-pressure modification of CdVO$_3$. The high-pressure phase has a perovskite-type structure and exhibits metallic behavior [B. L. Chamberland and P. S. Danielson, J. Solid State Chem. \textbf{10}, 249 (1974)].} of CdVO$_3$. Its 1D crystal structure (Fig.~\ref{fig:structure}),\cite{onoda1999} featuring zigzag chains of edge-sharing VO$_5$ pyramids, seemingly represents an archetypal AFM insulator, where orbital degrees of freedom are eliminated by the square-pyramidal coordination of V$^{+4}$. The interchain couplings are long-range and likely AFM because of the underlying V--O--O--V superexchange pathways that typically lead to AFM couplings, while the FM superexchange is usually operative at short distances (see, e.g., Refs.~\onlinecite{pb2v3o9,nath2008,canadell2010}). Moreover, the next-nearest-neighbor intrachain coupling $J_2$ between the corner-sharing vanadium pyramids should also be AFM, as in CaV$_2$O$_5$ and related compounds (see Sec.~\ref{sec:discussion} and Table~\ref{tab:geometry}).\cite{korotin1999,korotin2000,pickett1997} However, experimental data disprove these empirical arguments, and reveal FM order in CdVO$_3$ below $T_C=24$~K.\cite{onoda1999} Magnetic susceptibility fitted with an expression for the classical FM chain model yields an effective intrachain coupling $J=-100$~K.\cite{onoda1999} Dai \textit{et al.}\cite{dai2003} found $J_1\simeq -288$~K and $J_2\simeq -90$~K from generalized gradient approximation (GGA) band structure calculations. They tentatively ascribed the unusual FM $J_2$ to the shift of the vanadium atom toward the base of the pyramid. However, their estimate of $J_1$ severely exceeds the experimental coupling of $-100$~K and, more importantly, does not address the nature of the interchain couplings, which are the driving force of the puzzling FM order in CdVO$_3$.

In the following, we explore the microscopic mechanism of ferromagnetism in CdVO$_3$ by extensive band structure calculations combined with magnetization measurements and quantum Monte-Carlo (QMC) simulations. The application of diverse approaches for the evaluation of exchange couplings, along with the direct comparison to the experimental data, leads to a reliable microscopic magnetic model of CdVO$_3$. We show that the FM behavior of this compound is peculiar, and relate the ferromagnetism to Cd $5s$ orbitals mediating the FM superexchange. Below, the methodological part (Sec.~\ref{sec:methods}) is followed by band structure results in Sec.~\ref{sec:band} and an analysis of the experimental data in Sec.~\ref{sec:experiment}. We conclude with a discussion and summary in Sec.~\ref{sec:discussion}.
\section{Methods}
\label{sec:methods}
The band structure was calculated within the framework of DFT using the full-potential local-orbital scheme (\texttt{FPLO8.50-32}).\cite{fplo} We applied the local-density-approximation (LDA) with the Perdew-Wang parametrization\cite{pw92} for the exchange-correlation potential. Exchange couplings were evaluated via two different procedures, a model approach and a supercell approach. In the former, the LDA band structure was mapped onto a multi-orbital Hubbard model that was further treated perturbatively in the strongly correlated limit. In the supercell approach, the correlation effects in the V $3d$ shell were treated in a mean-field fashion by the LSDA+$U$ method. Total energies for a number of collinear spin configurations were mapped onto a classical Heisenberg model to yield individual exchange couplings. Since the supercell approach led to somewhat puzzling results, we performed an extensive cross-check using: i) GGA\cite{pbe96} within \texttt{FPLO}; ii) Vienna \emph{ab initio} simulation package (\texttt{VASP})\cite{vasp1,*vasp2} that performs projected augmented wave calculations and therefore employs a different basis set.\cite{paw1,*paw2}
\footnote{The energy cutoff was set to 400~eV.}
LDA results were obtained for the orthorhombic crystallographic unit cell, with a fine $k$ mesh of 1040~points in the symmetry-irreducible part of the first Brillouin zone. DFT+$U$ calculations utilized supercells doubled along $b$ or $c$, with $k$ meshes of $150-200$~points.

To evaluate the magnetic susceptibility and Curie temperature of the proposed spin model, we performed QMC simulations using the \texttt{loop} algorithm\cite{loop} from the ALPS package.\cite{alps} 1D and two-dimensional (2D) finite lattices comprised $N=120$ and 512~($32\times 16$) sites, respectively, and ensured the absence of finite-size effects for the magnetic susceptibility within the temperature range under investigation. The Curie temperature was estimated from simulations for a three-dimensional (3D) model with different lattice sizes (see Sec.~\ref{sec:experiment}).

Experimental magnetization data were collected with a SQUID magnetometer (Quantum Design MPMS) in the temperature range $2-380$~K in applied fields up to 5~T. The single-phase polycrystalline sample of CdVO$_3$ was prepared by a solid-state reaction of CdO and VO$_2$ in an evacuated silica tube at 700~$^{\circ}$C for 24~hours. A 25~\% excess of CdO was introduced to compensate for the losses caused by the volatalization and reaction with the tube. The phase purity of the sample was checked by x-ray powder diffraction (Huber G670 Guinier camera, CuK$_{\alpha1}$ radiation, image-plate detector, $2\theta=3-100^{\circ}$ angle range).
\section{Band structure and exchange couplings}
\label{sec:band}
\subsection{LDA and model approach}
\begin{figure}
\includegraphics{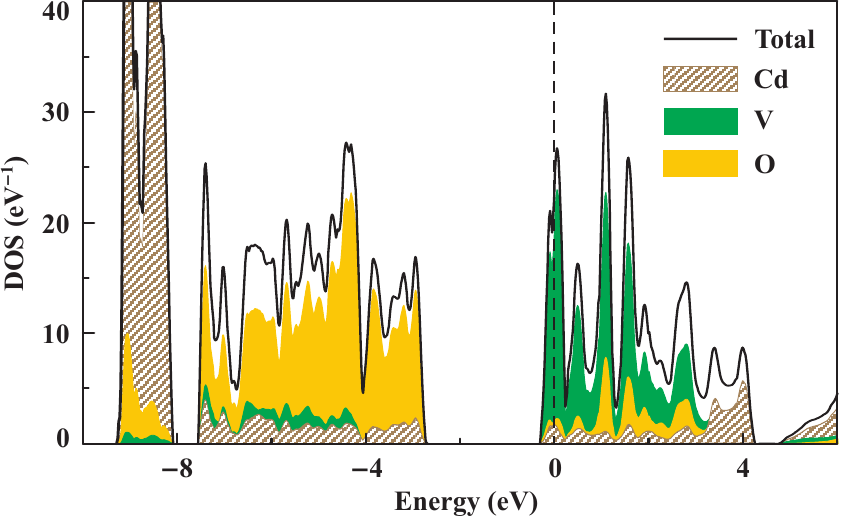}
\caption{\label{fig:dos}
(Color online) LDA density of states for CdVO$_3$. The Fermi level is at zero energy.
}
\end{figure}

The LDA band structure of CdVO$_3$ (Fig.~\ref{fig:dos}), with oxygen $2p$ valence bands below $-3$~eV and vanadium $3d$ bands at the Fermi level ($E_F$), is reminiscent of other V$^{+4}$ oxides.\cite{korotin2000,kaul2003} The contribution of cadmium is, however, larger than typical for an alkaline-earth (e.g., Ca$^{+2}$, Sr$^{+2}$)\cite{korotin2000,kaul2003} or even a $d^{10}$ (e.g., Ag$^{+1}$, Zn$^{+2}$)\cite{tsirlin2008,tsirlin2010} cation. The bands below $-8$~eV originate from the filled Cd $4d$ orbitals, whereas the states at $3-4$~eV show predominantly Cd $5s$ character. It is worth noting that the contributions of Cd and O at the Fermi level are comparable (6.2\% and 9.6\%, respectively, for $E\leq 0.2$~eV), yet oxygen $2p$ states dominate over Cd $5s$ between 1~eV and 3~eV. The sizable contribution at $E_F$ distinguishes Cd from other cations that also produce conduction bands $3-4$~eV above the Fermi level, but show a negligible contribution at $E_F$ [for instance, Pb$^{+2}$ in PbZnVO(PO$_4)_2$ (Ref.~\onlinecite{tsirlin2010}) or Se$^{+4}$ in VOSeO$_3$ (Ref.~\onlinecite{vose2o5})]. The obtained gapless energy spectrum originates from the underestimation of correlation effects in LDA. LSDA+$U$ reproduces a band gap of about 2.0~eV ($U_d=4$~eV, FPLO) in reasonable agreement with the brown color of CdVO$_3$.

\begin{figure}
\includegraphics{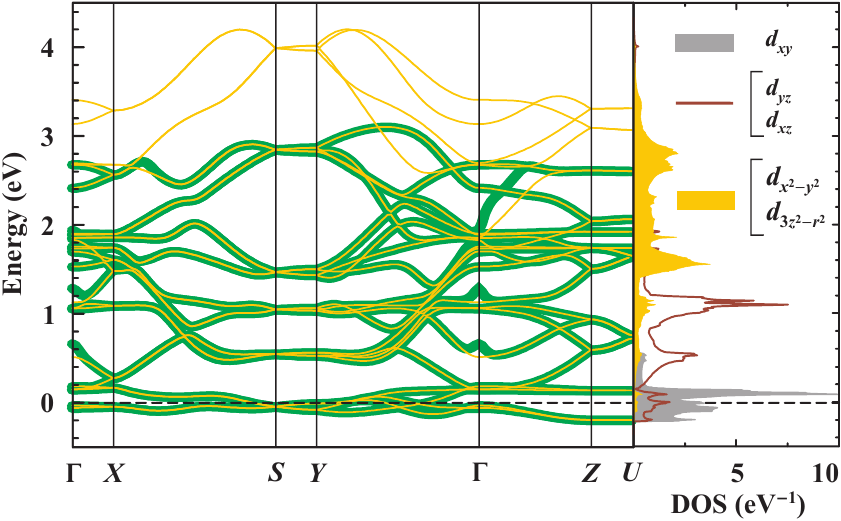}
\caption{\label{fig:band}
(Color online) Left: LDA band structure for CdVO$_3$ (thin light lines) and the fit of the tight-binding model (thick dark lines) for the V $3d$-related bands. Right: orbital-resolved DOS for V $3d$ states. The Fermi level is at zero energy. The poor fit in the vicinity of the $\Gamma$ point is related to the strong hybridization with the Cd $5s$ states represented by the four high-lying bands (not fitted). The notation of $k$ points is: $\Gamma(0,0,0)$, $X(0.5,0,0)$, $S(0.5,0.5,0)$, $Y(0,0.5,0)$, $Z(0,0,0.5)$, and $U(0.5,0,0.5)$.
}
\end{figure}

The band complex between $-0.5$~eV and 4.2~eV comprises 24 bands (Fig.~\ref{fig:band}). Since there are four formula units per cell, these bands arise from six orbitals per formula unit: five V $3d$ and one Cd $5s$. The $3d$ levels of vanadium lie below the Cd states, and show a crystal-field splitting characteristic of V$^{+4}$O$_5$ square pyramids.\cite{tsirlin2009,valenti2003} Directing the $z$ axis along the short (apical) V--O bond of the VO$_5$ pyramid, we find the lowest-lying $d_{xy}$ crystal-field level, represented by four narrow bands at the Fermi level (Fig.~\ref{fig:band}). 

To analyze exchange couplings, we fit 20 vanadium bands%
\footnote{The fit is complicated by the substantial hybridization between V $3d$ and Cd $5s$ states. Since the ambiguity of the fit could affect the hopping parameters and thus the resulting exchange integrals, we carefully tested our estimates by varying energy windows for Wannier functions or adding explicit contributions of Cd $5s$ orbitals as ``tails''. We also constructed a 24-band model including Wannier functions with Cd $5s$ character. All these tests produced rather similar results and confirmed our main conclusion on the FM superexchange in CdVO$_3$.} with a tight-binding model based on Wannier functions adapted to specific orbital symmetries,\cite{wannier} and map the resulting transfer integrals $t$ (hoppings)\cite{supplement} onto a multi-orbital Hubbard model with the effective on-site Coulomb repulsion $U_{\eff}=4$~eV and Hund's coupling $J_{\eff}=1$~eV.\cite{agvaso5,pb2v3o9} In the $t\ll U_{\eff}$ limit, this model is reduced to the Kugel-Khomskii model, and the exchange couplings are expressed as follows:\cite{kugel,mazurenko2006}
\begin{equation}
  J=\dfrac{4t_{xy}^2}{U_{\eff}}- \sum_{\alpha}\dfrac{4t_{xy\rightarrow\alpha}^2J_{\eff}}{(U_{\eff}+\Delta_{\alpha})(U_{\eff}+\Delta_{\alpha}-J_{\eff})},
\label{eq:exchange}\end{equation}
where $t_{xy}$ and $t_{xy\rightarrow\alpha}$ are transfers between the $xy$ states and from the $xy$ (half-filled) to $\alpha$ (empty) states, respectively.\cite{supplement} $\Delta_{\alpha}$ stands for the crystal-field splitting between the $xy$ and $\alpha$ states. Note that the states with a certain $3d$ orbital character and the resulting hoppings in Eq.~\eqref{eq:exchange} refer to Wannier functions centered on vanadium sites. Each Wannier function entails one of the vanadium $3d$ orbitals along with oxygen $2p$ and cadmium $5s$ states (Fig.~\ref{fig:wannier}). Therefore, metal--ligand transfers are implicitly contained in the hopping parameters $t_i$. 

With Eq.~\eqref{eq:exchange}, one evaluates full exchange couplings that are a sum of the AFM superexchange, arising from the transfers between the half-filled $xy$ states [Eq.~\eqref{eq:exchange}, first term], and the FM superexchange due to the hoppings to empty $d$ states [Eq.~\eqref{eq:exchange}, second term]. The efficiency of the LDA-based model approach has been demonstrated in Refs.~\onlinecite{tsirlin2010,agvaso5,pb2v3o9} providing a direct comparison to the experiment. 

\begin{table}
\caption{\label{tab:model}
Interatomic distances (in~\r A) in the CdVO$_3$ structure and the exchange couplings (in~K) calculated with Eq.~\eqref{eq:exchange}: the AFM ($J^{\AFM}$) and FM ($J^{\FM}$) contributions and the resulting total exchange ($J$).
}
\begin{ruledtabular}
\begin{tabular}{ccccr}
       & Distance & $J^{\AFM}$ & $J^{\FM}$ & $J$   \\
 $J_1$ & 3.05     & 18         & $-87$     & $-69$ \\
 $J_2$ & 3.60     & 21         & $-30$     & $-9$  \\
 $J_3$ & 5.93     & 0          & $-8$      & $-8$  \\
 $J_c$ & 5.20     & 0          & $-17$     & $-17$ \\
 $J_a$ & 5.79     & 1          & $-4$      & $-3$  \\
\end{tabular}
\end{ruledtabular}
\end{table}
The leading exchange couplings calculated according to Eq.~\eqref{eq:exchange} are listed in Table~\ref{tab:model}. The FM components of $J_1$ and $J_2$ surpass the AFM superexchange. The interchain coupling along $c$ lacks any AFM component, yet there is a sizable FM contribution of $-17$~K. The interchain coupling along $a$ is weaker but also FM. Further couplings are below 2~K (in terms of the absolute value) with an exception of $J_3\simeq -8$~K, which is the third-neighbor intrachain coupling. 

The exclusively FM couplings in CdVO$_3$ readily lead to the FM long-range order. Therefore, the results of the model approach are consistent with the experimental data, at least on the qualitative level. In Sec.~\ref{sec:experiment}, we will further demonstrate a good quantitative agreement with the experiment, while the rest of the present section is focused on the application of the supercell approach (Sec.~\ref{sec:dft+u} and~\ref{sec:comparison}) and on the microscopic origin of ferromagnetism in CdVO$_3$ (Sec.~\ref{sec:ferro}).

\subsection{DFT+$U$ puzzles}
\label{sec:dft+u}
The supercell approach to the evaluation of exchange couplings is based on LSDA+$U$ (more generally, DFT+$U$) calculations that effectively reproduce the gapped energy spectrum of a Mott insulator. The DFT+$U$ method rests upon the mean-field solution of the Hubbard model in the strongly correlated limit, but the incorporation of this solution into the self-consistent procedure leads to several features making the DFT+$U$ results different from that of the perturbative treatment on top of LDA (model approach). One difference is the application of the on-site Coulomb repulsion and exchange to individual atomic-like orbitals in DFT+$U$ instead of hybridized LDA bands (corresponding to molecular-like orbitals or Wannier functions) in the model approach. Therefore, the DFT+$U$ parameters $U_d$ and $J_d$ are generally different from $U_{\eff}$ and $J_{\eff}$ in the Hubbard model.\cite{cu2v2o7,[{See also: }][{}]mazurenko2007,*mazurenko2008} The second feature is the double-counting-correction (DCC) scheme that subtracts part of the Coulomb energy already contained in LSDA, and enables the self-consistent procedure. Depending on the filling of individual $d$ orbitals, the DCC is applied in either around-mean-field (AMF) or fully-localized-limit (FLL) fashions. 

The choices of $U_d$, $J_d$, and the DCC are made on empirical grounds, and retain a certain ambiguity. Here, we focus on the previously overlooked effect of DCC, with $U_d$ and $J_d$ fixed at 3~eV and 1~eV, respectively. These values have been justified by \texttt{FPLO} LSDA+$U$ calculations for several V$^{+4}$ compounds.\cite{tsirlin2009,pb2v3o9,agvaso5} We also varied $U_d$ in the physically reasonable range of $2.5-6$~eV, but no qualitative differences were found. To avoid calculations for large supercells, we estimated $J_1+J_3$ instead of evaluating $J_1$ and $J_3$ separately. According to Table~\ref{tab:model}, $|J_3|\!\ll |J_1|$, i.e., one may assume $J_1+J_3\simeq J_1$.

\begin{table}
\caption{\label{tab:lsda+u}
Exchange couplings (in~K) evaluated by the DFT+$U$ supercell procedure for different functionals, double-counting-correction (DCC) schemes, and band structure codes. The DFT+$U$ parameters are set to $U_d=3$~eV (\texttt{FPLO}), $U_d=4$~eV (\texttt{VASP}), and $J_d=1$~eV (both codes).
}
\begin{ruledtabular}
\begin{tabular}{rrrrccc}\medskip
$J_1+J_3$ & $J_2$ & $J_a$ & $J_c$ & Functional & DCC & Code          \\
  $-17$ & 21    &   2   &   9   &  LSDA+$U$  & AMF & \texttt{FPLO} \\\smallskip
  $-18$ & 30    &  2    &   8   &  GGA+$U$   & AMF & \texttt{FPLO} \\
 $-130$ & 11    &  $-4$ & $-24$ &  LSDA+$U$  & FLL & \texttt{FPLO} \\\smallskip
 $-117$ & 17    &  $-3$ & $-17$ &  GGA+$U$   & FLL & \texttt{FPLO} \\
 $-122$ & $-4$  &  $-3$ & $-26$ &  LSDA+$U$  & FLL & \texttt{VASP} \\
\end{tabular}
\end{ruledtabular}
\end{table}
The AMF and FLL results for the exchange couplings in CdVO$_3$ are rather different (Table~\ref{tab:lsda+u}).\cite{supplement} AMF evaluates the mostly antiferromagnetic scenario, while FLL renders $J_1$ strongly FM. We further checked these results against the different exchange-correlation potential (GGA+$U$) and the different band structure code. For a given DCC, the LSDA+$U$ and GGA+$U$ estimates closely match. The \texttt{VASP} calculations can be done for FLL only, and support the respective estimates from \texttt{FPLO}. A marginal difference between the \texttt{FPLO} and \texttt{VASP} results (especially for the second-neighbor coupling $J_2$) is related to different basis sets and, consequently, different projection schemes employed in the construction of the DFT+$U$ occupation matrix. This difference also explains the 1~eV offset in the value of $U_d$ (see Ref.~\onlinecite{cu2v2o7}). Another remark regards the GGA (without $U$) results by Dai \textit{et al.}\cite{dai2003} who found both $J_1$ and $J_2$ FM, with a clearly overestimated absolute value of $J_1$. Their results cannot be directly compared to ours, because uncorrelated GGA calculations heavily underestimate correlation effects (for example, the reported band gap of about 0.5~eV\cite{dai2003} is much too small to explain the brown color of CdVO$_3$). The neglect of strong correlation effects typically leads to huge errors in the exchange couplings, as demonstrated, e.g., in Ref.~\onlinecite{pbvo3}.

Table~\ref{tab:lsda+u} shows that the DFT+$U$ estimates are robust with respect to the exchange-correlation potential and to the particular basis set employed in the band structure code. Thus, the problem stems from the choice of the DCC, which has a strong and unanticipated effect on the computed exchange couplings. A simple qualitative analysis identified the correct ground state for the FLL set of exchange parameters, whereas AMF predicts the wrong ground state. Indeed, the FM ground state in CdVO$_3$ requires FM interchain couplings $J_a$ and $J_c$ and the FM or weakly AFM $J_2$. The AFM next-nearest-neighbor intrachain coupling $J_2$ frustrates $J_1$, but does not break the FM ground state for $J_2/|J_1|<\frac14$ (Ref.~\onlinecite{[{For example: }][{}]zinke2009}). The FLL results $J_a,J_c<0$ and $J_2/|J_1|\leq 0.15$ fulfill both conditions. By contrast, $J_2/|J_1|\simeq 1$ obtained in AMF induces a spiral order along the chains that are further coupled antiferromagnetically. Therefore, the AMF-based scenario is unrealistic.

The above analysis puts forward the advantages of FLL in evaluating the exchange couplings for CdVO$_3$. This conclusion is reinforced by the physical meaning of different DCC schemes.\cite{ylvisaker2009} While AMF describes the regime of moderate correlations, FLL should be appropriate for strongly correlated systems ($t_i\ll U_{\eff})$, such as CdVO$_3$. To check whether this DCC can be used universally, we calculated exchange couplings for several simple V$^{+4}$ compounds and further explored the effect of the DCC on the stability of different magnetic states in DFT+$U$.

\subsection{Double-counting correction: AMF vs. FLL}
\label{sec:comparison}
To compare the performance of the AMF and FLL versions of DFT+$U$, we select five representative V$^{+4}$ compounds showing chain-like magnetic behavior (Table~\ref{tab:comparison}). While details of the interchain couplings and the magnetic ground state may be different (and, in some cases, not fully understood), the AFM nature of the intrachain couplings is safely established experimentally.\cite{mgvo3,vosb2o4,kaul2003,csvof3,voac2003} The key difference between our test systems is the mutual arrangement of the vanadium polyhedra. To analyze the connectivity, we only consider the magnetic unit, a VO$_5$ square pyramid, despite the actual crystallographic local environment is often described as an octahedron. The advantage of our description is the simple identification of the relevant superexchange pathway through the oxygen atoms lying in the basal plane of the pyramid. Since all systems under consideration reveal the $d_{xy}$ orbital ground state, the axial oxygen atom neither connects the pyramids nor contributes to the superexchange, and only the oxygen atoms in the basal plane take part in the superexchange couplings. 

\begin{figure}
\includegraphics{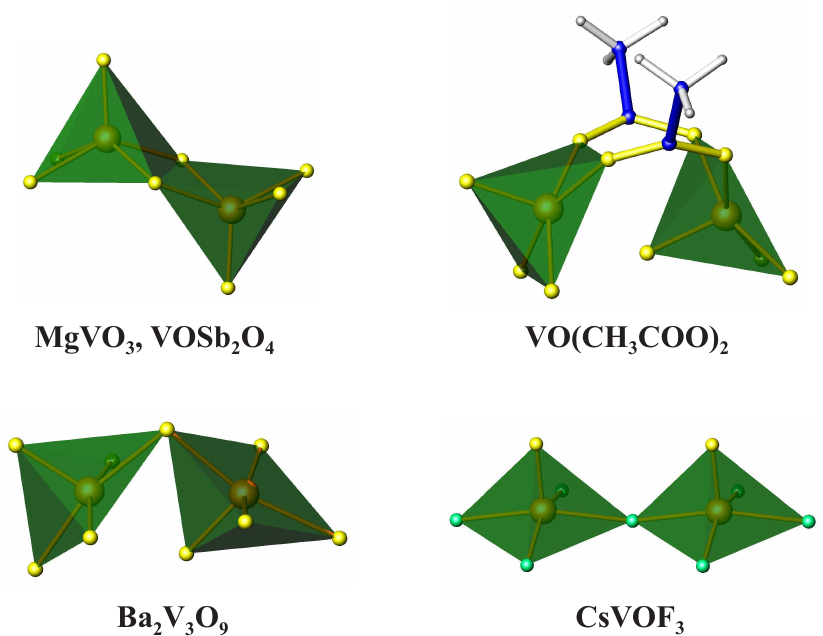}
\caption{\label{fig:test}
(Color online) Superexchange pathways in the test compounds listed in Table~\ref{tab:comparison}.
}
\end{figure}

The test systems represent several different regimes (Fig.~\ref{fig:test}): i) edge-sharing VO$_5$ pyramids (VOSb$_2$O$_4$ and the low-pressure modification of MgVO$_3$);\cite{mgvo3,vosb2o4} ii) corner-sharing VO$_5$ pyramids with either twisted (Ba$_2$V$_3$O$_9$, V--O--V angle of $96.4^{\circ}$)\cite{kaul2003} or nearly linear (CsVOF$_3$, V--O--V angle of $164.9^{\circ}$)\cite{csvof3} geometries; iii) VO$_5$ pyramids bridged by a non-magnetic acetate (CH$_3$COO$^-$) group.\cite{voac2003} 

The intrachain exchange coupling $J$ is evaluated as the energy difference between the FM and AFM spin configurations. Further couplings, both interchain and long-range intrachain, are rather weak and irrelevant for the present analysis. Quantitative estimates of these weak couplings and detailed structural information can be found in the preceding studies, where the exchange integrals were successfully evaluated using the model approach.\cite{chaplygin1999,chaplygin2004,kaul2003,koo2010} 

\begin{table}
\caption{\label{tab:comparison}
Test quasi-1D V$^{+4}$ compounds, the V--V distances $d$ (in~\r A), the connections between the VO$_5$ pyramids, and the exchange couplings $J$ (in~K) obtained from the AMF or FLL supercell calculations and from the experiment.
}
\begin{ruledtabular}
\begin{tabular}{ccccccc}
\medskip
  Compound & $d(\text{V--V})$ & Connection & $J^{\AMF}$ & $J^{\FLL}$ & $J^{\exp}$ & Ref. \\
  MgVO$_3$       &  2.98  & edge-sharing   & 128 & $-84$  & 100 & \onlinecite{mgvo3}    \\
  VOSb$_2$O$_4$  &  3.01  & edge-sharing   & 248 &  154   & 250 & \onlinecite{vosb2o4}  \\
  Ba$_2$V$_3$O$_9$ & 3.01 & corner-sharing &  79 &   84   &  94 & \onlinecite{kaul2003} \\
  CsVOF$_3$      &  3.91  & corner-sharing & 143 &  157   & 132 & \onlinecite{csvof3}   \\
  VO(CH$_3$COO)$_2$     & 3.48   & via ac bridges & 544 &  528   & 430 & \onlinecite{voac2003} \\
\end{tabular}
\end{ruledtabular}
\end{table}

\begin{figure}
\includegraphics{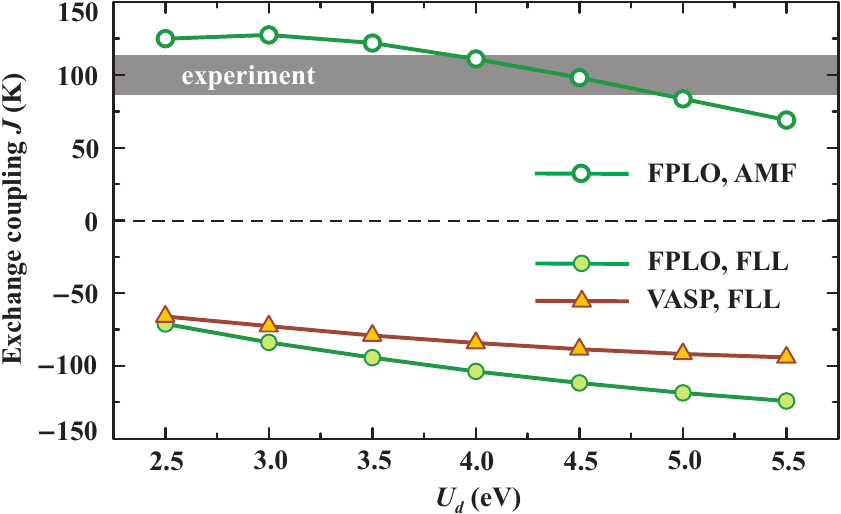}
\caption{\label{fig:mgvo3}
(Color online) Intrachain exchange coupling in MgVO$_3$ calculated by LSDA+$U$ for different values of Coulomb repulsion parameter $U_d$ and different DCC schemes: AMF (open symbols) and FLL (filled symbols). The marginal difference between the FLL results obtained in \texttt{FPLO} and in \texttt{VASP} is likely related to the different basis sets.
}
\end{figure}

In Table~\ref{tab:comparison}, we compare AMF and FLL results for the leading intrachain exchange couplings.%
\footnote{Here, we use $U_d=3$~eV for five-fold-coordinated V$^{+4}$ (MgVO$_3$, VOSb$_2$O$_4$), $U_d=4$~eV for six-fold-coordinated V$^{+4}$ (Ba$_2$V$_3$O$_9$, CsVOF$_3$, VO(ac)$_2$), and $J_d=1$~eV for all test compounds. The octahedral oxygen environment requires a higher $U_d$, as shown in, e.g., Refs.~\onlinecite{tsirlin2008,tsirlin2009}.} The corner-sharing (Ba$_2$V$_3$O$_9$, CsVOF$_3$) or indirect [VO(CH$_3$COO)$_2$] connections between the VO$_5$ pyramids lead to a remarkably small difference between AMF and FLL. By contrast, there are large discrepancies for the couplings between edge-sharing pyramids, especially in MgVO$_3$. These discrepancies are basically independent of the $U_d$ value, because $J^{\AMF}$ and $J^{\FLL}$ show similar evolution with a nearly constant offset of $J^{\AMF}-J^{\FLL}\simeq 100$~K in VOSb$_2$O$_4$ and about 200~K in MgVO$_3$ (Fig.~\ref{fig:mgvo3}). In VOSb$_2$O$_4$, the discrepancy might be still tolerable, because both types of DCC yield the correct antiferromagnetic solution. The weaker exchange coupling, along with a larger offset between AMF and FLL, render the FLL result for MgVO$_3$ qualitatively wrong. Fig.~\ref{fig:mgvo3} shows that any reasonable value of $U_d$ leads to negative $J^{\FLL}$ violating the experimental AFM coupling. The \texttt{VASP} calculations for MgVO$_3$ produce similar $J$ values and confirm the intrinsic nature of the problem. 

Based on our findings for the test compounds, we arrive at two conclusions that are -- at this stage -- rather empirical: i) the particular choice of the DCC scheme is relevant for short-range couplings only; more specifically, only the couplings between the edge-sharing vanadium pyramids are affected; ii) for edge-sharing pyramids, neither DCC can be used universally, because FLL produces a realistic scenario for CdVO$_3$, while failing for MgVO$_3$, and in AMF the situation is exactly the opposite. We believe that the broad range of coupling geometries considered in our study makes this empirical conclusions a helpful guidance for future computational work on V$^{+4}$ oxides and other transition-metal compounds. We also emphasize the remarkably good performance of the model approach for CdVO$_3$. The model approach is free from the double-counting problem, lacks any ambiguity, and represents an appealing alternative to the DFT+$U$ supercell calculations. A further discussion of methodological aspects and tentative remarks on the possible origin of the DCC effects are given in Sec.~\ref{sec:discussion}.

\subsection{Origin of ferromagnetism}
\label{sec:ferro}
The FM behavior of CdVO$_3$ contrasts with the AFM properties of other V$^{+4}$ compounds, such as quasi-1D MgVO$_3$ and VOSb$_2$O$_4$ or quasi-2D CaV$_2$O$_5$, CaV$_3$O$_7$, and CaV$_4$O$_9$. While $J_1$ is a nearly $90^{\circ}$ and possibly FM V--O--V superexchange, the weakness of $J_2$ as well as the FM nature of $J_a$ and $J_c$ are less clear from a microscopic point of view. Dai \textit{et al.}\cite{dai2003} claimed that the shift of the vanadium atom toward the basal plane of the VO$_5$ pyramid could cause FM $J_2$. However, they did not verify this conjecture, and refrained from any analysis of the FM interchain couplings, which are crucial for the FM long-range order. Here, we argue that the origin of ferromagnetism in CdVO$_3$ is different, and relates to the admixture of Cd states to the magnetic orbitals (Wannier functions). We justify this mechanism by considering a model system, a hypothetical CaVO$_3$ compound with the crystal structure of CdVO$_3$ but Ca occupying the Cd position. In our analysis, we follow the approach of Refs.~\onlinecite{korotin2000,tsirlin2009,tsirlin2010} that utilize the ability of DFT to evaluate microscopic parameters of fictitious compounds and, therefore, investigate the influence of different structural features on the magnetic properties.

\begin{figure}
\includegraphics{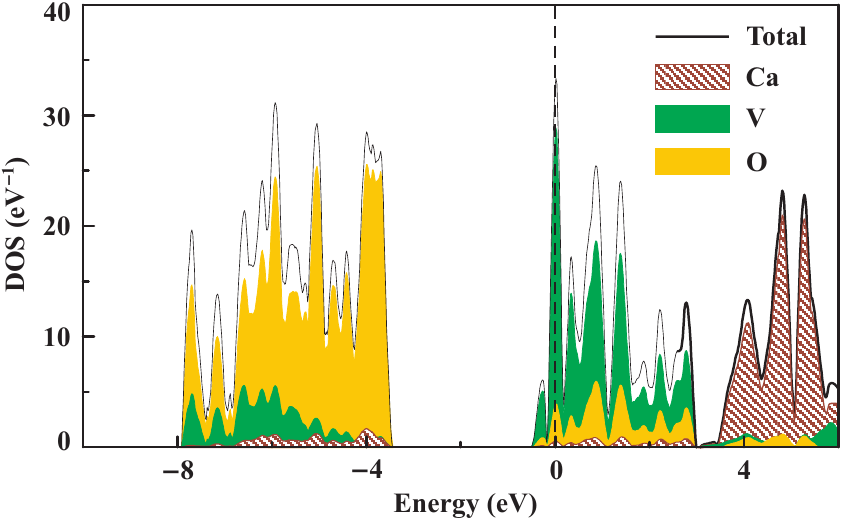}
\caption{\label{fig:ca}
(Color online) LDA density of states for the fictitious CdVO$_3$-type CaVO$_3$ (compare to Fig.~\ref{fig:dos}). The Fermi level is at zero energy. 
}
\end{figure}
The band structures of CdVO$_3$ and hypothetical CaVO$_3$ are rather similar, yet CaVO$_3$ is free from the low-lying Cd $5s$ bands (compare Figs.~\ref{fig:dos} and~\ref{fig:ca}). Therefore, vanadium bands are largely mixing with O $2p$ and basically lack the cation contribution.\cite{supplement} This change in the band structure has a strong effect on individual transfer integrals and the resulting exchange couplings (Table~\ref{tab:ca}). Compared to Table~\ref{tab:model}, we find: i) the increase in AFM $J_2$; ii) the AFM nature of $J_a$ and $J_c$; iii) isotropic interchain couplings ($J_a\simeq J_c$). The hypothetical CaVO$_3$ is predicted to be predominantly antiferromagnetic, and conforms to the trends established for V$^{+4}$ compounds. Specifically, the long-range interchain couplings are AFM, whereas $J_2$ is about 100~K, as in MgV$_2$O$_5$ and CaV$_4$O$_9$ (see Sec.~\ref{sec:discussion} and Table~\ref{tab:geometry}). 

\begin{table}
\caption{\label{tab:ca}
Exchange couplings (in~K) in the hypothetical CaVO$_3$ compound with the CdVO$_3$ structure. The $J^{\AFM}$, $J^{\FM}$, and $J=J^{\AFM}+J^{\FM}$ values are obtained from Eq.~\eqref{eq:exchange} (model approach), whereas the numbers in the last two columns (AMF and FLL) are calculated via the supercell approach with different DCC (LSDA+$U$, $U_d=3$~eV, $J_d=1$~eV). Similar to Table~\ref{tab:lsda+u}, the LSDA+$U$ estimates of $J_1$ are in fact $J_1+J_3$ with $J_3\simeq 5$~K according to the model analysis.
}
\begin{ruledtabular}
\begin{tabular}{c@{\hspace{3em}}ccc@{\hspace{3em}}cc}
       & \multicolumn{3}{c}{Model approach} &  \multicolumn{2}{c}{LSDA+$U$} \\
       &  $J^{\AFM}$ & $J^{\FM}$ &  $J$     & $J$        & $J$        \\
       &             &           &          &  AMF       & FLL        \\\hline
 $J_1$ &     26      &   $-42$   & $-16$    & $-28$      & $-163$     \\
 $J_2$ &    105      &   $-23$   &   82     & 133        & 83         \\
 $J_c$ &      0      &       0   &   0      & 1          & $-3$       \\
 $J_a$ &      1      &    $-1$   &   0      & 2          & $-2$       \\
\end{tabular}
\end{ruledtabular}
\end{table}

The LSDA+$U$ results for CaVO$_3$ reveal the same strong dependence on the DCC, as previously observed in CdVO$_3$ (compare Tables~\ref{tab:lsda+u} and~\ref{tab:ca}). The AMF and FLL calculations basically agree on the sizable AFM $J_2$, but strikingly differ in the estimate of the short-range coupling $J_1$. While we can not obtain any experimental information on the hypothetical CaVO$_3$ compound, it is instructive to compare the LSDA+$U$ estimates of $J_1$ to the independent result from the model approach. The model estimate of $J_1\simeq -16$~K is now in good agreement with the AMF prediction $J_1\simeq -28$~K, while the FLL estimate of $J_1\simeq -163$~K is far too large in terms of the absolute value. A further discussion of methodological problems related to the DCC of DFT+$U$ is given in Sec.~\ref{sec:discussion}.

\begin{figure}
\includegraphics{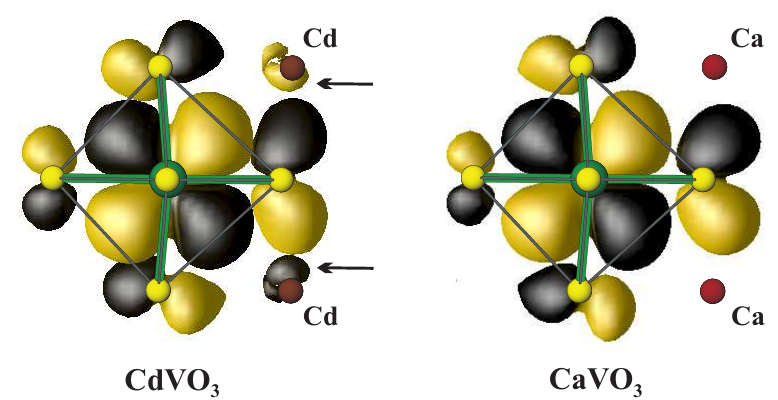}
\caption{\label{fig:wannier}
(Color online) Vanadium $d_{xy}$-based Wannier functions for CdVO$_3$ (left) and hypothetical CaVO$_3$ (right). Note the Cd $5s$ contributions (marked with arrows) that are missing in the Ca-containing compound.
}
\end{figure}
The role of the Cd $5s$ states in the magnetism of CdVO$_3$ is elucidated by Wannier functions. In Fig.~\ref{fig:wannier}, we compare the V $d_{xy}$-based Wannier functions for CdVO$_3$ and CaVO$_3$. Apart from the leading V $d_{xy}$ contributions, both Wannier functions involve oxygen $p$ orbitals. In CdVO$_3$, there is an additional Cd $5s$ contribution, which is missing in CaVO$_3$. Similar features are found for Wannier functions based on empty vanadium orbitals. The Cd orbitals represent ``tails'' of the Wannier functions and amplify interorbital hoppings that drive the FM superexchange. 
\section{Analysis of the experimental data}
\label{sec:experiment}
\begin{figure}
\includegraphics{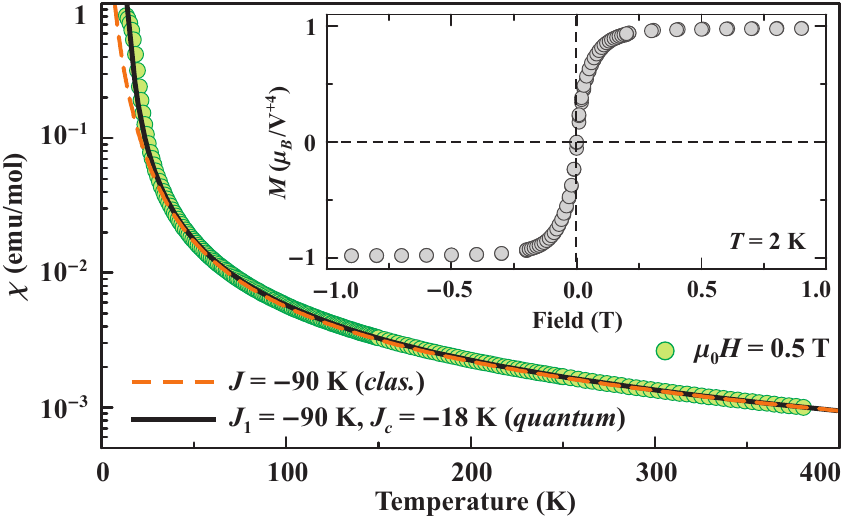}
\caption{\label{fig:chi+mvsh}
(Color online) Magnetic susceptibility of CdVO$_3$ measured in an applied field of 0.5~T (filled circles) and the fits with the classical 1D model (dashed line) as well as the quantum 2D $J_1-J_c$ model (solid line). Inset: magnetization curve at 2~K.
}
\end{figure}

The magnetic susceptibility ($\chi$) of CdVO$_3$ steeply increases from 380~K to 2~K and clearly indicates the FM nature of the system (Fig.~\ref{fig:chi+mvsh}). At low temperatures, the magnetization reaches the saturated value of about 1~$\mu_B$ at the applied field of less than 0.2~T. The lack of hysteresis may be related to a very weak anisotropy of V$^{+4}$ (Ref.~\onlinecite{[{For example: }][{}]gnezdilov2008}). Above 230~K, the susceptibility follows the Curie-Weiss law 
\begin{equation}
  \chi=\chi_0+\dfrac{C}{T+\theta},
\end{equation}
where $\chi_0=-70\times 10^{-6}$~emu/mol is the temperature-independent contribution of core diamagnetism and van Vleck paramagnetism, $C=0.368$~emu~K~mol$^{-1}$ is the Curie constant, and $\theta=-46$~K is the Weiss temperature. The $C$ value corresponds to the effective magnetic moment of 1.70~$\mu_B$ that perfectly matches the expected value of $g\mu_B\sqrt{S(S+1)}=1.697$~$\mu_B$ with $g=1.96$ from ESR.\cite{onoda1999} The negative Weiss temperature is a signature of FM couplings leading to a positive deviation from the Curie-Weiss behavior below 230~K. 

\begin{figure}
\includegraphics{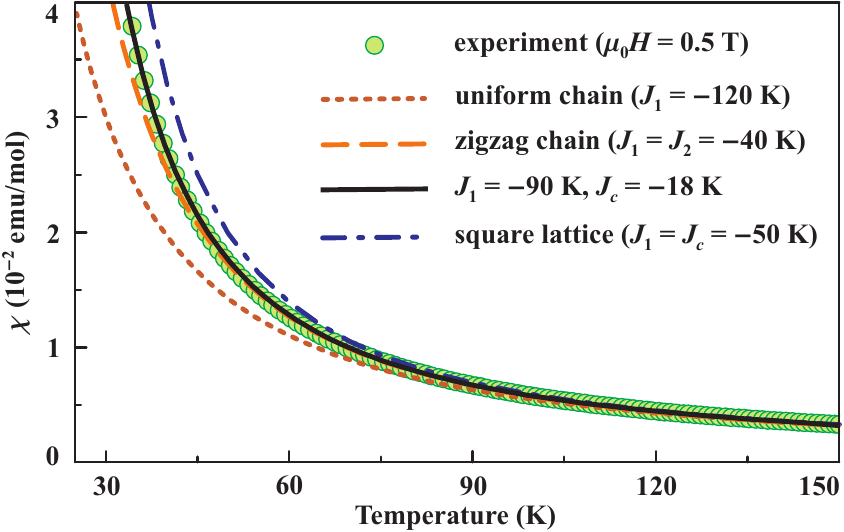}
\caption{\label{fig:fits}
(Color online) Fits of the magnetic susceptibility with different 1D and 2D spin models, see text for details.
}
\end{figure}

Onoda and Nishiguchi\cite{onoda1999} fitted the magnetic susceptibility with the expression for the classical spin chain:\cite{fisher1964}
\begin{equation}
  \chi=\chi_0+\dfrac{N_Ag^2\mu_B^2}{4k_BT}\dfrac{1+u}{1-u},\quad u=\text{coth}\left(-\dfrac{3J}{4T}\right)+\dfrac{4T}{3J}.
\label{eq:classical}\end{equation}
Our data can be described in a similar way ($\chi_0=-70\times 10^{-6}$~emu/mol, $g=1.99$, $J=-90$~K, dashed line in Fig.~\ref{fig:chi+mvsh}), but the model itself does not apply to the spin-$\frac12$ compound CdVO$_3$ because of inherent quantum fluctuations in low-dimensional spin-$\frac12$ systems. For example, at $S=\frac12$ the classical AFM chain shows the susceptibility maximum at $T_{\max}/J\simeq 0.35$, whereas for the quantum AFM chain $T_{\max}/J\simeq 0.64$. A pronounced difference should also be expected for the FM case. Indeed, the simulated curve for the quantum FM chain ($J=-120$~K) fits our experimental data down to 100~K only (short-dashed line in Fig.~\ref{fig:fits}). At lower temperatures, the 1D quantum model underestimates the susceptibility. This should be understood as an effect of quantum fluctuations that disturb the parallel alignment of spins and therefore reduce $\chi$. 

Since the experimental data for CdVO$_3$ conform to the classical model (see Fig.~\ref{fig:chi+mvsh}), quantum fluctuations in this compound are less pronounced than in the single quantum spin chain. This reduction could be related to the interchain coupling $J_c$ that increases the dimensionality, or the FM intrachain coupling $J_2$ that increases the number of bonds at a lattice site without changing the dimensionality. We start with the first option, which is also favored by DFT since $|J_2|\ll |J_1|$. The 2D $J_1-J_c$ model fits the experimental data down to 30~K with $J_1=-90$~K, $J_c/J_1=0.2$, $g=1.96$, and $\chi_0=-80\times 10^{-6}$~emu/mol (solid line in Fig.~\ref{fig:fits}). For comparison, we also considered the isotropic 2D model (FM square lattice, $J_c=J_1=-50$~K) where quantum fluctuations are further suppressed by the increased dimensionality. This model consequently overestimates the susceptibility below 90~K (dash-dotted line in the same figure). Therefore, CdVO$_3$ exhibits an intermediate regime between 1D and 2D FM systems.

The long-range FM order in CdVO$_3$ is stabilized by the interchain couplings $J_a$ and $J_c$. To evaluate the Curie temperature $T_C$, we considered the 3D $J_1-J_c-J_a$ spin model and calculated the Binder ratio of magnetization $B(T)=\langle m^4\rangle/\langle m^2\rangle^2$ for finite lattices of different size. The Curie temperature was determined as the crossing point of several $B(T)$ curves calculated for $L\times L/2\times L/2$ finite lattices with $L\leq 64$. Similar to Refs.~\onlinecite{sandvik1999,sengupta2009}, we reduced the dimensions of the lattices along $J_a$ and $J_c$ to account for the anisotropic nature of our system. Using $J_a/J_1=0.03$ (Tables~\ref{tab:model} and~\ref{tab:lsda+u}), we arrive at $T_C/J_1=0.212$ ($T_C=19$~K) that is slightly below the experimental value of 24~K.%
\footnote{To reach the experimental Curie temperature, one has to take $J_a/J_1=0.07$ which is, however, inconsistent with the susceptibility fit. Since there are four couplings $J_a$ and two coupling $J_c$ at each lattice site, $J_a/J_1=0.07$ increases the overall energy of the interchain couplings by 70~\%, and alters the susceptibility fit.}

The marginal underestimate of $T_C$ in the $J_1-J_a-J_c$ model may be related to magnetic anisotropy, which lies beyond the scope of the present study, or the second-neighbor intrachain coupling $J_2$. According to DFT results, this coupling is either weakly FM (Table~\ref{tab:model}) or weakly AFM (Table~\ref{tab:lsda+u}). The AFM $J_2$ introduces frustration and disables the QMC techniques because of the sign problem. By contrast, the effect of FM $J_2$ can be readily evaluated. At weak FM $J_2$, the overall energy of the intrachain exchange ($-90$~K) is merely redistributed between $J_1$ and $J_2$. Using $J_1=J_2=-40$~K, we are able to introduce more significant changes by reducing quantum fluctuations and improving the susceptibility fit even within the purely 1D $J_1-J_2$ model (long-dashed line in Fig.~\ref{fig:fits}). The second-neighbor coupling $J_2$ brings the system closer to the classical regime, thereby reducing the fitted interchain coupling. Since the ordering temperature is mostly sensitive to the value of $J_c$, the $J_1-J_2-J_a-J_c$ model further underestimates $T_C$. Therefore, the sizable FM $J_2$ is unlikely. In conclusion, we argue that the unfrustrated $J_1-J_a-J_c$ spin lattice, providing a remarkable fit of the susceptibility and a reasonable estimate of $T_C$, is a valid microscopic model of CdVO$_3$. This model strongly supports our computational results: compare the experimental $J_1\simeq -90$~K, $J_a\simeq -3$~K, and $J_c\simeq -18$~K to the calculated $J_1\simeq -69$~K, $J_a\simeq -3$~K, and $J_c\simeq -17$~K (Table~\ref{tab:model}). 
\section{Discussion and summary}
\label{sec:discussion}
The peculiar ferromagnetism of CdVO$_3$ puts forward several important issues. First, we have demonstrated the importance of the DCC scheme as one of the delicate parameters of the DFT+$U$ method. While it is usually difficult to make a well-justified choice of the DCC scheme, empirical recipes can be used, as we show below. We found that FLL produces accurate estimates for certain compounds, such as CdVO$_3$, but it may fail in closely related systems (e.g., MgVO$_3$) where AMF is, on the contrary, the method-of-choice. The most pragmatic and safe solution to this problem would be the adjustment of the DCC against the experimental data for each specific compound. While this does not diminish the crucial role of band structure calculations in the microscopic modeling of complex materials, the constant reference to the experimental data deprives computational methods of one important advantage, the ability to predict the magnetism of hitherto unexplored systems. To remedy this drawback, we performed a test study of several V$^{+4}$ compounds with different coupling geometries.

Based on the comparative study in Sec.~\ref{sec:comparison}, we conclude that the ambiguity related to the choice of DCC is confined to short-range couplings only. Moreover, the uncertainty is only present for edge-sharing VO$_5$ pyramids in MgVO$_3$, CdVO$_3$, and VOSb$_2$O$_4$. The similarly short V--V distance of about 3.0~\r A in Ba$_2$V$_3$O$_9$ with the corner-sharing pyramids does not cause any problems: both AMF and FLL results are in excellent agreement with the experiment (Table~\ref{tab:comparison}). The principal difference between the edge-sharing and corner-sharing geometries is the combination of the direct V--V exchange and \mbox{V--O--V} superexchange in the former, while only the \mbox{V--O--V} superexchange is featured by the latter. It is the combination of the direct exchange and superexchange or the direct exchange itself that are not well reproduced by DFT+$U$, presumably, due to the oversimplified mean-field treatment of correlation effects. The accurate results of the LDA-based model approach (Table~\ref{tab:model}) ensure the high accuracy of the hopping parameters used as an input. Therefore, the underlying interatomic interactions are well reproduced even in LDA, but a better treatment of the on-site correlation effects is required. The development of the respective computational techniques will be a rewarding but challenging task that lies far beyond the scope of our study.

Presently, we are able to formulate the following recipes for calculating exchange couplings in transition-metal compounds. The conventional supercell approach (DFT+$U$ calculations for different spin configurations) can be safely used for long-range couplings and even for short-range couplings not involving the direct overlap of the magnetic orbitals. In this case, the choice of the DCC plays a minor role compared to other factors, such as the adjustment of the Coulomb repulsion parameter $U_d$. Once the direct overlap of the magnetic orbitals is encountered, DFT+$U$ results should be taken with caution and tested against experimental data or against independent computational estimates. Particularly, we put forward the model approach as an appealing alternative to the DFT+$U$ calculations. The perturbative treatment of the multi-orbital Hubbard model is based on the reliable LDA input, free from ambiguity, and uncovers individual hopping processes that are responsible for the superexchange. We believe that the combination of the model and supercell approaches is a viable and reliable tool for studying magnetic systems, as demonstrated by our present work on CdVO$_3$. 

After commenting on the methodological aspects, we discuss the physics of CdVO$_3$. The unexpected FM behavior of this compound originates from Cd $5s$ states mediating the hoppings between the half-filled and empty states of V$^{+4}$. The interaction is, therefore, a superexchange, but its mechanism is different from the conventional orbital ordering scenario.\cite{kugel,khomskii1973} The orbital ordering induces magnetic (half-filled) orbitals of different symmetry on the neighboring atoms, and favors the hoppings between the half-filled and empty orbitals of the same symmetry. In CdVO$_3$, there is only one type of the magnetic orbital, hence ferromagnetism is driven by hoppings between orbitals of different symmetry. 

\begin{table}
\caption{\label{tab:geometry}
  Exchange couplings between corner-sharing VO$_5$ pyramids: V--O--V angles (in~deg), V--V distances (in~\r A), and exchange couplings $J$ (in~K) estimated from the experiment (marked with an asterisk) or DFT calculations.
}
\begin{ruledtabular}\medskip
\begin{tabular}{cccrc}
  Compound         & $\alpha$(V--O--V) & $d$(V--V) & $J$     & Ref.                     \\
  Ba$_2$V$_3$O$_9$ &   96.4            &  3.01     & 94$^*$  & \onlinecite{kaul2003}    \\
  MgV$_2$O$_5$     &  117.6            &  3.37     & 92      &                          \\
                   &  141.1            &  3.69     & 144     & \onlinecite{korotin1999} \\
  CaV$_4$O$_9$     &  129.9            &  3.54     & 148     & \onlinecite{korotin1999} \\
  CaV$_2$O$_5$     &  132.9            &  3.49     & 608     &                          \\
                   &  135.3            &  3.60     & 122     & \onlinecite{korotin1999} \\
  CdVO$_3$         &  136.1            &  3.60     & $<20$   & This work                \\
  PbVO$_3$         &  147.7            &  3.80     & 203$^*$ & \onlinecite{tsirlin2008} \\
  CsVOF$_3$        &  164.9            &  3.91     & 132$^*$ & \onlinecite{csvof3}      \\
\end{tabular}
\end{ruledtabular}
\end{table}
The effect of Cd is easily recognized in Table~\ref{tab:geometry} that compares exchange couplings between the corner-sharing VO$_5$ pyramids. Although one generally expects the increase in the AFM superexchange at V--O--V angles close to $180^{\circ}$, this trend does not hold for V$^{+4}$ compounds due to the relevance of other structural features. For example, basal planes of the pyramids coincide in CaV$_2$O$_5$ and CsVOF$_3$, but remain nearly perpendicular in Ba$_2$V$_3$O$_9$ (Fig.~\ref{fig:test}). This explains the broad range of possible V--O--V angles and their weak effect on the superexchange. The only general trend is the AFM nature of the coupling between the corner-sharing VO$_5$ pyramids. The couplings typically range between 100 and 200~K, and always exceed 90~K. CdVO$_3$ is a remarkable exception, with the AFM coupling reduced below 20~K because of the Cd $5s$ orbitals altering the superexchange. 

The cation-mediated superexchange in vanadium oxides is not restricted to CdVO$_3$. For example, the unusually strong and frustrating second-neighbor exchange in PbVO$_3$ is likely driven by Pb $6p$ orbitals that marginally contribute to the V $d_{xy}$ bands.\cite{pbvo3}
\footnote{To verify the role of Pb in PbVO$_3$, we evaluated the second-neighbor coupling $J_2$ in the fictitious SrVO$_3$ having the tetragonal PbVO$_3$ structure. The calculations yield $J_2=24$~K in SrVO$_3$ against $J_2=69$~K in PbVO$_3$.} A similar ``diagonal'' superexchange has been proposed for Pb$_{0.55}$Cd$_{0.45}$V$_2$O$_5$, although the intrinsic disorder of Pb and Cd atoms in this compound hampered an accurate experimental estimate of the respective coupling.\cite{tsirlin2007} The aforementioned systems are, however, different from CdVO$_3$, because the Pb and Cd cations mediate an AFM superexchange. The non-magnetic cations may have diverse effects on the superexchange, and the specific scenario is actually determined by the hybridization between the cation orbitals and oxygen orbitals entering vanadium-based Wannier functions. In many systems (e.g., vanadium phosphates\cite{tsirlin2009}), there is no effect at all because the cation states are expelled from the Fermi level. The cation-mediated superexchange is not ubiquitous and requires specific coupling geometries, but it can be expected for a variety of cations featuring empty $s$ (e.g., Zn$^{+2}$, Cu$^{+1}$, Ag$^{+1}$) or $p$ (e.g., Bi$^{+3}$, Sn$^{+2}$) states closely above the Fermi level (see also Ref.~\onlinecite{geertsma1996}). Respective compounds are likely to host non-trivial spin lattices and unusual magnetism.

Finally, we note that CdVO$_3$ is interesting on its own as a system showing an intermediate regime between the 1D and 2D ferromagnets. Experimental studies of FM uniform chains helped to find appropriate theoretical tools for solving the Heisenberg model in 1D, and disclosed peculiar soliton-type excitations.\cite{kopinga1989,*campana1990,*kopinga1993,kopinga1987,*devries1989} The respective systems are close to the 1D limit, whereas the opposite, 2D limit is realized in K$_2$CuF$_4$ and related Cu$^{+2}$ halides with layered perovskite-type structures.\cite{yamada1972,*ito1976,khomskii1973,dejongh1976,*willett1988} CdVO$_3$ could be a reference point between these two qualitatively different model regimes that represent the ground states with zero (1D) and non-zero (2D) ordered magnetic moment. The crossover between the 1D and 2D regimes is rather well studied for the AFM case,\cite{sandvik1999} yet a comparative study of the FM case could be insightful.

In summary, we have developed a microscopic magnetic model of CdVO$_3$, and found out the origin of ferromagnetism in this compound. We argue that CdVO$_3$ is a ferromagnetic spin chain system with an effective (and mostly nearest-neighbor) intrachain coupling of $-90$~K and a sizable interchain coupling $J_c=-18$~K along one of the dimensions. Our model is based on extensive band structure calculations and verified by a direct comparison to the experimental magnetic susceptibility and Curie temperature. The unusual ferromagnetic couplings arise from Cd $5s$ orbitals that contribute to the vanadium-based Wannier functions, mediate hoppings between the half-filled and empty $d$ states of vanadium, and lead to the ferromagnetic superexchange. This mechanism puts forward diverse effects of non-magnetic cations on superexchange in transition-metal compounds.

\acknowledgments
We are grateful to Yurii Prots and Horst Borrmann for x-ray diffraction measurements. A.T. was supported by Alexander von Humboldt foundation.

%

\end{document}